\documentclass[aps,reprint, floatfix,superscriptaddress,twocolumn,showpacs,amsmath,amssymb]{revtex4-2}
\usepackage{graphicx} 
\usepackage{color}
\usepackage[number-unit-product=~, per-mode=symbol, separate-uncertainty = true, range-units=single, range-phrase=\text{--}, list-units=single, list-final-separator={\text{, and }}]{siunitx}
\newcommand{\percentage}[1]{\num{#1}\si{\percent}}
\newcommand{\SIhyphen}[2]{\num{#1}-\si{#2}}
\usepackage[version=4]{mhchem}
\usepackage{physics}
\usepackage{mathtools}

\usepackage[utf8]{inputenc}
\usepackage[T1]{fontenc}
\usepackage{times}
\usepackage{etoolbox}
\newcommand{\subm}[1]{_{\mathrm {#1}}}

\newcommand{\rev}[1]{{#1}}
\newcommand{\sub}[1]{\ensuremath{_{\mathrm{#1}}}}

\newcommand{\phim}{\phi\subm{m}}
\newcommand{\thetaK}{\theta\subm{K}}

\newcommand{\thetaT}{\theta\subm{tot}}

\begin{document}

\title{
Magneto-optical Kerr-effect measurements under pulsed magnetic fields over 40~T \\ using a compact sample fixture}

\author{Atsutoshi~Ikeda}
\email{ikeda.atsutoshi.3w@kyoto-u.ac.jp}
\affiliation{Department of Electronic Science and Engineering, Kyoto University, Kyoto 615-8510, Japan}

\author{$^{\!,\,\dagger}$\ Sota~Nakamura}
\thanks{These authors contributed equally to this work.}
\affiliation{Department of Electronic Science and Engineering, Kyoto University, Kyoto 615-8510, Japan}

\author{Soichiro~Yamane}
\affiliation{Department of Electronic Science and Engineering, Kyoto University, Kyoto 615-8510, Japan}

\author{Kosuke~Noda}
\affiliation{Department of Engineering Science, University of Electro-Communications, Tokyo 182-8585, Japan}

\author{Akihiko Ikeda}
\affiliation{Department of Engineering Science, University of Electro-Communications, Tokyo 182-8585, Japan}

\author{Shingo~Yonezawa}
\email{yonezawa.shingo.3m@kyoto-u.ac.jp}
\affiliation{Department of Electronic Science and Engineering, Kyoto University, Kyoto 615-8510, Japan}

\date{\today}

\begin{abstract}
\rev{The} magnetic field is one of the most fundamental control parameters in materials science.
\rev{A} pulsed magnetic-field apparatus can generate high magnetic fields that are inaccessible by conventional DC-field magnets.
One important issue is that measurement techniques compatible with pulsed fields are rather limited due to short pulse duration and large electromagnetic or mechanical \rev{noise} originating from field pulses.
\rev{The} magneto-optical Kerr effect (MOKE), \rev{the} change in \rev{the state} of light polarization upon reflection from magnetic materials, has \rev{the} potential to become a powerful tool for investigation of magnetic properties of a wide range of materials including non-transparent materials or thin films in pulsed fields.
Nevertheless, since \rev{the} MOKE response is typically very small, MOKE measurements under pulsed fields are quite challenging.
Here, we present a new method to measure polar MOKE under high pulsed magnetic fields of 2-ms pulse width.
The keys of this new technique are a ferrule-based compact sample-fiber fixture and a phase-resolved numerical lock-in analysis, combined with the high-resolution optical apparatus based on an all-fiber loop-less Sagnac interferometer.
We succeeded in measuring MOKE \rev{signals from} various ferromagnetic or ferrimagnetic samples above 40~T and down to 77~K, \rev{significantly} extending \rev{the} limits of previously reported pulse-field MOKE measurements.
Our apparatus is simple enough to be compatible with larger-scale experiments in pulse-field facilities, thus becoming a new promising tool to optically investigate \rev{material} properties in pulsed fields. 


\end{abstract}

\maketitle



\section{Introduction}\label{sec:intro}

\rev{The} magnetic field is a fundamental control parameter in materials science.
Under very high magnetic fields, the Zeeman energy experienced by an electron spin becomes comparable to the thermal energy.
In such high fields, typically larger than tens of tesla, new phases of matter can be induced.
Indeed, several field-induced phase transitions \rev{suggesting} such unconventional mechanisms have been observed, e.g., the insulator--metal transition in \ce{VO2}~\cite{Matsuda2020.NatureCommun.11.3591}, \rev{the} structural transition in solid oxygen~\cite{Nomura2014.PhysRevLett.112.247201}, and exciton condensation in \ce{LaCoO3}~\cite{Ikeda2023.NatureCommun.14.1744}.
Therefore, strong magnetic \rev{fields open} a new frontier in materials research.

\rev{A} pulsed magnetic \rev{field is} a powerful tool to induce and investigate such high-field phenomena.
While the \rev{world record for the highest} static magnetic field is \SI{45.5}{\tesla}~\cite{Hahn2019.Nature.570.496}, the pulsed field  in large facilities \rev{currently} reaches even \SI{1200}{\tesla}~\cite{Nakamura2018.RevSciInstrum.89.095106}.
\rev{In addition}, in ordinary laboratories, a portable pulse-magnet system \rev{can now} create fields larger than \SI{40}{\tesla}~\cite{Ikeda2024.JApplPhys.136.175902}, which is comparable to the \rev{world's} highest static field.
An important challenge \rev{in} pulse-field \rev{experiments} is that the typical duration of the pulsed field is of the order of \si{\milli\second} for a \SIhyphen{100}{\tesla} pulse and \si{\micro\second} for a \SIhyphen{1000}{\tesla} pulse.
Such short \rev{durations} require measurements at frequencies of the order of \si{\mega\hertz} and thus \rev{restrict the range of} applicable techniques.

High measurement frequencies above \si{\mega\hertz} can naturally be achieved by optical probes.
When the light interacts with a material without the time-reversal symmetry, \rev{right-} and left circularly polarized \rev{light} gain different phases. 
The cause of this difference in \rev{phase} is called the magneto-optical (MO) effect. 
The MO effect can \rev{also be described}  as a \rev{rotation of the polarization of} linearly polarized light.
The MO effect seen via transmission is called the magneto-optical Faraday effect, and the \rev{effect observed} via reflection is \rev{called} the magneto-optical Kerr effect (MOKE).
The polarization rotation angle due to the Faraday effect, the Faraday angle $\theta\subm{F}$, can be as large as several radians \rev{for transparent samples with large thickness}, and thus the Faraday effect has \rev{already been} used in pulse-field experiments~\cite{Levitin2002.PhysSolidState.44.2107, Nakamura2018.RevSciInstrum.89.095106}.
In contrast, pulse-field MOKE measurements are much more challenging, since \rev{the} typical polarization rotation angle due to the Kerr effect, namely the Kerr angle $\thetaK$, is at most of the order of \rev{milliradians}. 
Indeed, there are only a few examples of pulse-field MOKE studies, where measurements \rev{have only been performed} up to 11~T and at room temperature~\cite{Chen2013.Measurement.46.52, Lin2017.ChinJPhys.55.698}

On the other hand, MOKE experiments have \rev{lead to} various innovations in recent materials science.
Fascinating examples include discoveries of chiral superconductivity accompanied by spontaneously broken time-reversal symmetry in the superconducting state of \ce{Sr2RuO4}~\cite{Xia2006.PhysRevLett.97.167002}, \ce{UPt3}~\cite{Schemm2014.Science.345.190}, and \ce{URu2Si2}~\cite{Schemm2015.PhysRevB.91.140506}.
The observation of the quantization of the Kerr angle in graphene~\cite{Shimano2013.NatureCommun.4.1841}, reminiscent \rev{of} the quantum Hall effect, demonstrates \rev{the} importance of the band topology in optical phenomena.
The discovery of the first van der Waals ferromagnet \ce{CrI3} was largely owing to MOKE measurements~\cite{Huang2017.Nature.546.270}.
These studies also demonstrate a general advantage of MOKE over the Faraday effect, namely, that it is applicable to thin samples such as films and two-dimensional materials as well as to non-transparent materials such as conductive materials.
Considering discoveries achieved by these previous MOKE studies and \rev{the} wide applicability \rev{of MOKE}, it is natural to expect that MOKE measurements in high pulsed fields, if realized, \rev{will yield further significant discoveries} in high-field regimes.

To solve difficulties in realizing pulse-field MOKE due to the smallness of the MOKE response, we need a high-resolution measurement technique.
Here, we came up with the idea to combine pulsed fields with the loop-less (zero-loop-area) Sagnac interferometer~\cite{Xia2006.ApplPhysLett.89.062508}, which has been used to detect \rev{subtle but non-trivial} spontaneous magnetic signals in zero or small static fields~\cite{Xia2006.PhysRevLett.97.167002, Schemm2014.Science.345.190, Schemm2015.PhysRevB.91.140506, Saykin2023.PhysRevLett.131.016901, 
Hu2023.arXiv2208.08036, Farhang2023.NatureCommun.14.5326, Wang2024.PhysRevMaterials.8.014202}.
\rev{However, the} very short time scale for measurements in pulsed fields urges us to develop a new analysis method.
In addition, we needed to develop a new compact and stable setup that fits inside a small bore of a pulse-field magnet \rev{and is capable of overcoming} electromagnetic and mechanical noises originating from the large current and Lorentz force \rev{generated} by the pulse-magnet system.

In this Letter, \rev{we report \rev{a} successful observation of polar MOKE up to \SI{43}{\tesla}, enabled by a new compact sample fixture for optical measurements combined with a portable pulse-field system and a phase-resolved numerical lock-in analysis}.
To the best of our knowledge, our results record the highest magnetic field under which the Kerr effect \rev{has ever been} observed.
The sample fixture utilizes standard connector parts of optical fibers and provides a plug-and-play measurement free from complicated alignment processes of the incident light.
Our setup, enabling quick magnetic measurements above \SI{40}{\tesla} in the laboratory, \rev{offers a new route to explore} magnetic properties of materials under extreme conditions.

\section{Ferrule-based compact sample fixture}

\rev{In optical measurements, including MOKE measurements, it is important to ensure a proper illumination of the sample and, further, to achieve optimal coupling of light reflected from the sample into the photodetector.}
In previous pulse-field MOKE studies, ordinary free-space optics \rev{were} used and sample coupling was achieved by precise alignment of the optical path using \rev{lenses} and mirrors~\cite{Chen2013.Measurement.46.52,Lin2017.ChinJPhys.55.698}.
Extensions of such free-space-based MOKE measurements to  higher fields or to lower temperatures are not straightforward.

Instead of free-space optics, we intend to use \rev{a} fiber-based loop-less Sagnac interferometer for pulse-field MOKE measurements.
In previous MOKE studies using the loop-less Sagnac interferometer in DC fields, various setups, such as miniature tripod structures, have been used, so that the light reflected from the sample is coupled back to the \rev{polarization-maintaining (PM)} fiber~\cite{Xia2008.PhDthesis, Hu2023.arXiv2208.08036,Wang2024.PhysRevMaterials.8.014202}.
Nevertheless, such setups are generally too large to be placed in small bores (with diameters typically less than $\phi 10$~mm) of pulse-field magnets~\cite{Nakamura2018.RevSciInstrum.89.095106,Ikeda2024.JApplPhys.136.175902}.
\rev{In addition, under} pulsed magnetic fields, \rev{a short and } strong field pulse can cause vibrations in the setup, which would disturb the optical coupling. 
In addition, because \rev{a} pulsed field generates Eddy currents in metallic parts, it is important to avoid highly conductive metals in the setup.
Thus, it was necessary to invent a new compact, stable, and non-metallic setup.

\begin{figure}[bt]
    \includegraphics[width=0.5\textwidth]{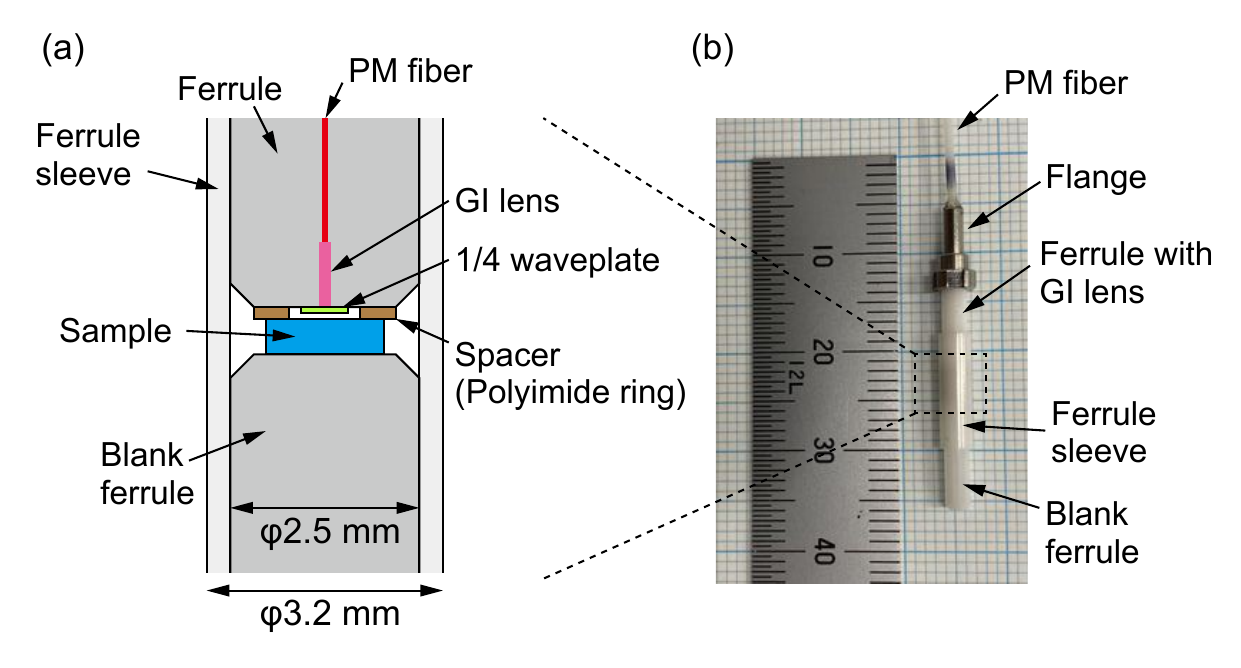}
    \caption{(a) Schematic of the central part of the sample fixture composed of a ferrule-embedded gradient-index (GI) lens focuser, 1/4 waveplate, ferrule sleeve, and polyimide spacer.
    The spacer thickness is chosen to be \SI{50}{\micro\meter}, \rev{such} that the sample surface is placed at the focal position of the GI lens simply by sandwiching the sample with two ferrules. 
    The diameter of the ferrules \rev{is} 2.5~mm and the outer diameter of the ferrule sleeve is \SI{3.2}{\milli\meter}.
    (b) Photo of the whole fixture. The ferrule focuser and the blank ferrule are fixed \rev{using} a ferrule sleeve.}
    \label{fig: fixture}
\end{figure}

In this work, we propose a new setup that is based on ceramic ferrules.
Ferrules are small rods \rev{that are} used in optical fiber connectors.
They are made with high \rev{precision, making them} suitable for optical coupling, although they are not costly. 
They are typically made with non-magnetic zirconia ceramics, which is \rev{suitable for avoiding} Eddy-current effects in pulsed magnetic fields.
As shown in Fig.~\ref{fig: fixture}, we used a \rev{custom} PM-fiber focuser made with \rev{a} gradient index (GI) lens embedded in a 2.5-mm diameter ceramic ferrule (Photonic Science Technology).
This focuser has \rev{a} focal length of \SI{50}{\micro\meter} and the \rev{diameter of the focused beam} is around \SI{6}{\micro\meter}.
\rev{A 1/4 waveplate made of thin polyimide} (NTT Advanced Technology Cooperation, AT-QWP-4A; $0.75\times 0.5\times 0.015$~mm$^3$) is \rev{glued at} the tip of the focuser.
The optical axes of the waveplate are 45$^\circ$ tilted with respect to the slow/fast axes of the PM fiber, so that the linearly polarized lights from the focuser are converted to the free-space circular polarized lights and vice versa.
The sample is placed beneath the focuser, with a 50-\si{\micro\meter} thick polyimide spacer. 
This spacer makes the focal point of the light match the sample surface.
The sample is fixed inside a ferrule sleeve (Thorlabs, ADAF1), sandwiched by the ferrule focuser and by a blank ferrule (Thorlabs, CF128-10).
One side of the sample should have good optical reflection, whereas the other side \rev{does not need to} be optically flat, but should be parallel to the other side.
If the surfaces are not perfectly parallel, it might be helpful to \rev{place} a \rev{suitable} cushion material between the sample and the blank ferrule.

Figure~\ref{fig: fixture}(b) shows a photo of the \rev{entire} ferrule-based fixture. 
The outer diameter of the ferrule sleeve is 3.2~mm.
The maximal diameter of the setup is about 4.5~mm due to the metallic flange made of stainless steel.
Since \rev{the} electrical conductivity of stainless steel is relatively small, we do not notice any Eddy-current effects.
The total length of the fixture is around 30~mm.
Notice that it is possible to further minimize the fixture by using ferrules with 1.25~mm diameter and by avoiding \rev{the} use of the metallic flange. 

\begin{figure*}[t]
    \includegraphics[width=0.8\textwidth]{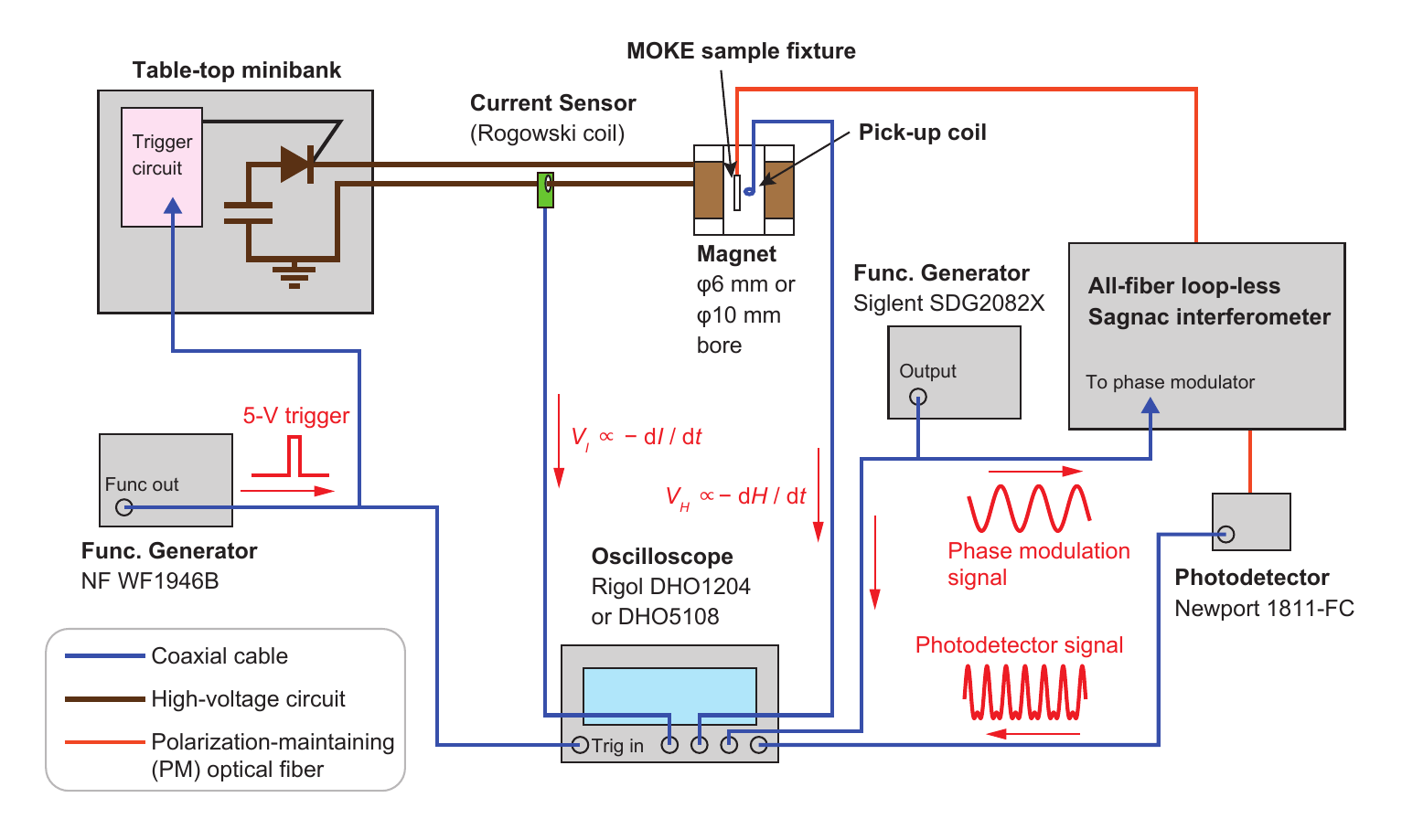}
    \caption{Measurement setup for the magneto-optical Kerr effect under pulsed magnetic \rev{fields}.
    Sets of voltage recorded by the oscilloscope are analyzed using the numerical lock-in analysis. When we use the 6-mm coil, the field pick-up coil and the Kerr fixture cannot be \rev{simultaneously} inserted to the bore. Thus we first run calibration with the pick-up coil in order to obtain the field-current relation. In subsequent MOKE measurements without the pick-up coil, the field value is calculated from the magnitude of the current.}
    \label{fig: setup}
\end{figure*}
 
This fixture \rev{demonstrates} very high and stable optical performance. 
We have found that more than 50\% of the incident light is coupled back to the fiber without any precise adjustments, if we use samples with relatively high reflectivity.
Thus, it \rev{only takes} a few minutes to prepare \rev{for the} experiments.
For comparison, it typically \rev{takes} more than one hour to optimize the optical coupling with the previous setup~\cite{Hu2023.arXiv2208.08036}.
This significant reduction of the preparation time is crucially important to achieve high-throughput experiments using pulse-field magnets, as well as other MOKE experiments using ordinary DC magnets.
Moreover, the optical coupling is quite stable, and moderate mechanical shocks do not affect the intensity of the reflected light.
We have also tested this fixture in low-temperature environments, such as \rev{through} direct \rev{immersion} in liquid nitrogen down to \SI{77}{\kelvin}, or dilute gas helium in a commercial cryostat (Quantum Design, PPMS) down to \SI{2}{\kelvin}.
In both cases, the change in the coupling is found to be typically around 10\%, as explained in Appendix~\ref{sec: appendix_T-dep-reflection}.

\section{Optic and electric apparatus}

For MOKE measurements, we used an all-fiber loop-less Sagnac interferometer. 
The basic concept of our interferometer is the same as those reported in Refs.~\cite{Xia2006.PhysRevLett.97.167002, Xia2006.ApplPhysLett.89.062508}.
The light source is a low-\rev{coherence} superluminescent diode (SLD) (Thorlabs, SLD1550P-A1 and SLD1550P-A40).
The \rev{emitted light has a center wavelength of} 1550~nm, corresponding to 0.8-eV photon energy.
The overall light intensity \rev{of} the interferometer was measured using an InGaAs photodetector (Newport, 1811-FC).
The main difference \rev{with respect to} Refs.~\cite{Xia2006.PhysRevLett.97.167002, Xia2006.ApplPhysLett.89.062508} is that we replaced free-space optical parts with PM-fiber-based optical instruments, such as a variable waveplate with a polarizer and a 45$^\circ$ \rev{principal}-axis splice (Phoenix, PWP-02-15-PM-2-1) and a phase modulator (iXblue Photonics, MPX-LN-0.1). 
This all-fiber setup has \rev{the advantage of being} easy to build and \rev{highly} stable against mechanical vibrations, etc. 
\rev{In addition}, fiber-based phase modulators operate in a much wider frequency range than free-space phase modulators. 

\rev{Figure \ref{fig: setup} illustrates our experimental setup.}
\rev{The} pulsed magnetic field was generated by a portable capacitor bank described in Ref.~\cite{Ikeda2024.JApplPhys.136.175902}. 
Our bank consists of \rev{a set of} dry-film 800-\SI{}{\micro\farad} 2-kV \rev{capacitors} (EPCOS, B25690C2807K003) and a 3-kV DC voltage source (Matsusada Precision, HUNS-3P117).
\rev{The circuit was designed to generate unipolar field pulses.}
A small coil with \rev{an} inner diameter of 10~mm or 6~mm, made with copper-silver alloy wire ($\phi 1.0$~mm, 14 turns $\times$ 10 layers) fixed with glass-powder-\rev{reinforced} epoxy, was connected to the capacitor bank.
The 10-mm coil can generate 12~T at room temperature with 900-V charging, and 34~T in liquid nitrogen with 1700-V charging; whereas the 6-mm coil can generate 15~T (800~V) at room temperature and 43~T (1700~V) at \SI{77}{\kelvin}.
The maximum field is limited by the Joule heating of the coil.
After charging the \rev{capacitors}, \rev{the} coil current is generated by activating the thyristor with a trigger pulse.
The circuit is shunted by a switch (not shown) after the generation of \rev{the} field pulse is completed.
\rev{The repetition rate of field pulses is limited by various factors.
For pulses close to the highest charging voltages mentioned above, it is safe to wait for approximately 10~min after each pulse to avoid accumulation of Joule heating.
For smaller pulses, we can achieve a repetition rate of around 2-3~min, which is limited by time consumption of other processes such as data saving, capacitor charging, etc. }

As shown in Fig.~\ref{fig: setup}, time-series data of the photodetector signal $V\subm{PD}(t)$, together with the current-sensor output, the magnetic-field pick-up coil output, and the phase-modulated signal from the function generator (Siglent, SDG2082X), were collected using a 12-bit digital oscilloscope (Rigol, DHO1204 or DHO5108).
Data acquisition was triggered by the 5-V pulse that was also used to trigger the output of the pulse-magnet current.
Data \rev{collected} for about \SI{10}{ms} with \rev{a} time interval of 1 or \SI{2}{\nano\second} amounted \rev{to} approximately 500~MB.


\section{Evaluation of magneto-optical rotation angle}
\label{sec: eval MO angle}

For ordinary steady-state MOKE measurements using \rev{the} loop-less Sagnac interferometer, lock-in amplifiers are used to extract the MOKE signal from the photodetector \rev{output}~\cite{Xia2006.PhysRevLett.97.167002,Xia2006.ApplPhysLett.89.062508, Saykin2023.PhysRevLett.131.016901, Farhang2023.NatureCommun.14.5326, Wang2024.PhysRevMaterials.8.014202}.
In contrast, for MOKE measurements under pulsed magnetic fields, it is difficult to use lock-in amplifiers, because their time constants \rev{must} be much shorter than the pulse width. 
\rev{In fact, we tried using} a digital lock-in amplifier (NF Corporation,  LI5660).
However, it turned out that the time constant had to be as short as \SI{1}{\micro\second} to avoid artificial delay of the response, \rev{while} the noise and higher-harmonics \rev{rejections were} not sufficient with such a short time constant.

Thus, we decided to use numerical lock-in analysis to obtain the MOKE signal from the time-dependent photodetector voltage signal.
In this work, we developed formulae to perform the numerical lock-in analysis with proper phase \rev{handling}. 
Analyses based on these equations enable us to determine both the amplitude and sign of the $n$-th harmonic signals in the photodetector voltage.
Details of this phase-resolved lock-in analysis \rev{are} described in Appendix~\ref{sec: appendix_numerical-lock-in} 1--3.

From the evaluated in-phase signals of the $n$-th harmonic signal $V_{\mathrm{in},n}$, we obtain the total MO rotation signal $\theta\subm{tot}$, which consists of the sample Kerr angle $\thetaK$ and background signal $\theta\subm{bk}$ originating from Faraday rotation of optical \rev{components}, using the relation
\begin{align}
   \theta\subm{tot} = \thetaK +\theta\subm{bk} \simeq \frac{1}{2}\frac{J_2(2\phim)}{J_1(2\phim)} \frac{V_{\mathrm{in},1}}{V_{\mathrm{in},2}}, 
   \label{eq:Kerr_conventional}
\end{align}
where $J_n(x)$ is the \rev{$n$-th order} Bessel function of the first kind.
For this relation, it is optimal to use the modulation depth $\phim$ of around $0.92$~rad, \rev{which corresponds to the maximum in $J_1(2\phim)$. }
For different modulation depths $\phim$, this relation can be generalized as 
\begin{align}
   \theta\subm{tot} \simeq \frac{1}{2}\frac{J_{m\subm{e}}(2\phim)}{J_{m\subm{o}}(2\phim)} \frac{V_{\mathrm{in},m\subm{o}}}{V_{\mathrm{in},m\subm{e}}},
   \label{eq:Kerr_fraction_genral}
\end{align}
where $m\subm{e}$ and $m\subm{o}$ are certain even and odd integers.
We can also use another convenient relation  
\begin{align}
    \theta\subm{tot} \simeq  \frac{V_{\mathrm{in},3}}{V_{\mathrm{in},2} + V_{\mathrm{in},4}} \sqrt{\frac{15V_{\mathrm{in},2} + 24 V_{\mathrm{in},4} + 9V_{\mathrm{in},6}}{80 V_{\mathrm{in},4}}},
    \label{eq: thetaK_new-relation}
\end{align}
when we use $\phim$ of around 2.1~rad (i.e., corresponding to the peak in $J_3(2\phim)$).
The proof of this relation is given in Appendix~\ref{subsec: evaluation_Kerr}.
This equation is \rev{useful} because one does not need to know $\phim$ very accurately before each measurement.
This advantage would be important in large-scale pulse-field experiments, where machine time is often quite limited.

A concrete example of \rev{the} data analysis process is described in detail in Appendix~\ref{sec: appendix_actual_process}.

\section{Results at Room Temperature}

\begin{figure}[t]
    \includegraphics[width=0.4\textwidth]{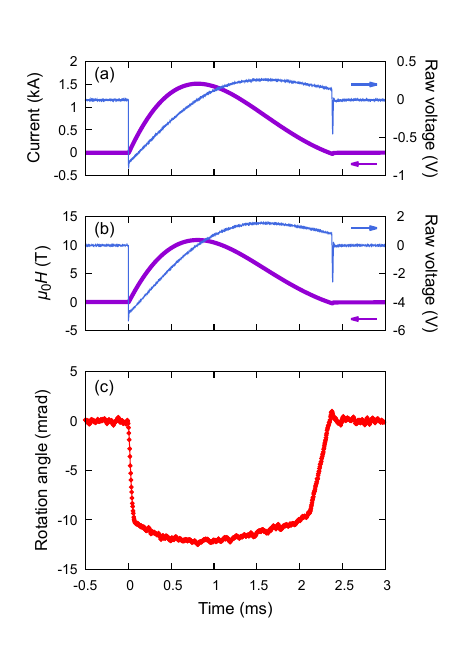}
    \caption{Typical voltage signals of (a) current and (b) field sensors recorded by the oscilloscope (blue curves; right vertical axes) for the 10-mm bore coil at room temperature. The current and field values evaluated by numerical integration of the sensor voltages are shown with purple curves (left vertical axes) in (a) and (b), respectively. (c) Time dependence of the magneto-optical rotation signal measured with Fe evaluated using numerical lock-in technique based on Eq.~\eqref{eq: thetaK_new-relation}.}
    \label{fig: pulse}
\end{figure}

In Fig.~\ref{fig: pulse}, we plot \rev{representative} time-dependent data obtained with the 10-mm bore coil at room temperature. 
The current and field sensor voltages are shown \rev{by} the blue curves in (a) and (b). 
Since these voltages are, respectively, proportional to $-dI/dt$ and $-dH/dt$, we can obtain coil current and magnetic field by numerically integrating the sensor voltages and then multiplying \rev{by appropriate} geometry factors. 
\rev{The resulting} $I(t)$ and $\mu_0H(t)$ curves are shown with the purple curves, demonstrating successful generation of \rev{a} field pulse with 2-ms duration.
\rev{The resolution of the magnetic-field sensor (pick-up coil) is 0.002-0.003 T, which is primarily attributed to electrical noise. 
In addition, the resolution is also limited by the temporal resolution (\SI{13}{\micro\second} for the data shown in this paper, as discussed in Appendix C). 
Within \SI{13}{\micro\second}, the magnetic field changes by 1\% of the pulse height in the beginning of the pulse.}

\begin{figure}[t]
    \includegraphics[width=0.4\textwidth]{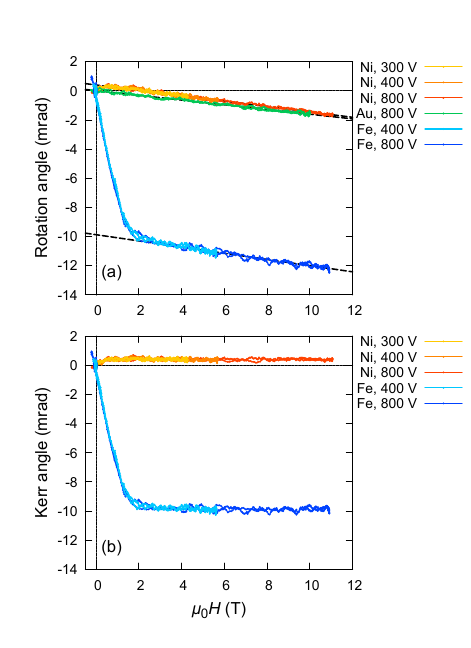}
    \caption{Results of pulse-field MOKE for various materials at room temperature plotted as a function of magnetic field.
    The optical power of the light source was \SI{0.8}{\milli\watt}.
    (a) Total measured MO rotation $\theta\subm{tot}$ containing the Kerr effect from the sample and the background Faraday effects. Here, $\thetaT$ was evaluated using Eq.~\eqref{eq: thetaK_new-relation}.
    (b) Kerr angle $\thetaK$ from the samples obtained by subtracting the field-linear background contribution.
    For both panels, curves with blue-like colors, orange-like colors, and green  represent data for Fe, Ni, and Au, respectively. 
    For Ni and Fe, data obtained with different charging voltages are shown.}
    \label{fig: Kerr}
\end{figure}

\rev{In the 2-ms pulse}, we succeeded in obtaining the MO rotation angle $\thetaT$ as shown in Fig.~\ref{fig: pulse}(c), by using the numerical lock-in analysis described in Sec.~\ref{sec: eval MO angle}.
\rev{
This curve was obtained using pure Fe (Nilaco, 99.99\% purity). 
When the pulse starts at $t=0$, a steep change in $\thetaT$ due to ferromagnetic domain alignment was observed.
Subsequently, a small and gradual modulation in $\thetaT$ was observed, which is attributed to the Faraday-effect background $\theta\subm{bk},$ as discussed in the next paragraph.  
Near the end of the pulse, a large change in $\thetaT$ was again observed, due to the randomization of ferromagnetic domains in the sample.
Since the coercivity of pure Fe is known to be very small, $\thetaT$ almost returns to zero when the field pulse finishes.
Notice that the field changing rate is not uniform within a pulse (See e.g., Fig.~\ref{fig: pulse}(b)).  
Since the field changes fastest in the beginning of the pulse, the change in $\thetaT$ is also fastest there; whereas the changes in $H$ and $\thetaT$ as a function of time is slower at the end of the pulse.
Because of the fast field change, the number of data points per magnetic field is lowest in the very beginning of the pulse, somewhat lowering the temporal and Kerr-angle resolutions. 
Nevertheless, this issue in measuring MOKE in low-field regimes can be simply overcome by repeating MOKE measurements with pulses of smaller heights, as demonstrated in Fig.~\ref{fig: Kerr}. 
}

Figure~\ref{fig: Kerr}(a) shows the field dependence of $\theta\sub{tot}$ for the ferromagnets \rev{Fe and} Ni (Nilaco, 99+\% purity), as well as non-magnetic Au (surface of gold-plated brass), \rev{measured with pulses of various heights}. 
For Fe, we polished the surface, whereas \rev{Ni and Au were used as bought without additional surface treatment.} 
\rev{Notice that the ``Fe, 800 V'' curve is plotted using exactly the same data as those of Fig.~\ref{fig: pulse}, without additional averaging or smoothing processes.
For clarity, data for Ni are also shown in Fig.~\ref{fig: Ni_kerr-vs-field} in Appendix \ref{sec: appendix_Ni} with a finer vertical scale.}
For Ni and Fe, steep changes in $\theta\sub{tot}$ in the low-field region due to saturation of the ferromagnetic moment were observed.
Under high fields above the saturation fields, all curves in Fig.~\ref{fig: Kerr}(a), including that of non-magnetic Au, exhibit linear field dependences with similar slopes of around \SI{-0.2}{\milli\radian\per\tesla}. 
Above saturation fields, the field dependence of $\thetaK$ \rev{of a ferromagnet is expected to} be very small~\cite{Visnovsky1993.JMagMagMat.127.135}.
Also, the intrinsic Kerr angle of Au is known to be \SI[retain-explicit-plus]{+0.0093}{\milli\radian\per\tesla}~\cite{Uba2017.PhysRevB.96.235132}, which is only 1/20 of the observed slope.
We have also found that the \rev{magnitude} of the slope changes when we intentionally change the position of the sample fixture inside the coil.
Thus, the linear field dependence commonly observed in the total rotation is a background contribution $\theta\subm{bk}$ and is attributable to the Faraday effect from the optical parts (lens, 1/4 waveplate, and fiber).
Notice that similar values of the slopes were obtained as long as samples were positioned exactly at the field center, as shown in Fig.~\ref{fig: Kerr}. 
This fact indicates the high reproducibility of the setup.

By subtracting $\theta\subm{bk}$ from $\theta\sub{tot}$, we obtain the Kerr angle shown in Fig.~\ref{fig: Kerr}(b).
For the background subtraction, we assume field-linear background $\theta\subm{bk} = \alpha \mu_0 H$ and fit the data under high field with the relation
$\theta\sub{tot}(H) = \alpha \mu_0 H+\theta\sub{K\infty}$, where $\theta\sub{K\infty}$ is the Kerr angle above the saturation field. 
\rev{Subsequently,} we subtracted the linear background term from the total rotation 
$\theta\sub{K}(H) = \theta\sub{tot}(H) - \alpha \mu_0 H$.
From this analysis, $\theta\sub{K\infty}$ was obtained to be $-9.87 \pm 0.03$~\si{\milli\radian} for Fe and {$+0.407 \pm 0.01$}~mrad for Ni.
Overall, the resolution of our experiment is of the order of 0.1~mrad, which is better than previous pulse-field MOKE studies (0.2-mrad resolution in Ref.~\cite{Chen2013.Measurement.46.52}).
The reported values of $\thetaK$ at 1550~nm wavelength \rev{fall within} the range of $-8.4~\text{mrad}\leq\thetaK\leq -7.6~\text{mrad}$ for Fe~\cite{Buschow1983.JMagMagMat.38.1, MacLaren1996.JApplPhys.79.6196, Delin1999.PhysRevB.60.14105} and within the range of $+0.31~\text{mrad}\leq \thetaK \leq +0.46~\text{mrad}$ for Ni~\cite{Buschow1983.JMagMagMat.38.1,Delin1999.PhysRevB.60.14105}.
Thus, our observed values of Fe are about 15-20\% larger than the reported values, while that of Ni is in \rev{the range of the variation in} the reported values.

\section{Results at Low Temperature}

\begin{figure}[t]
    \includegraphics[width=0.4\textwidth]{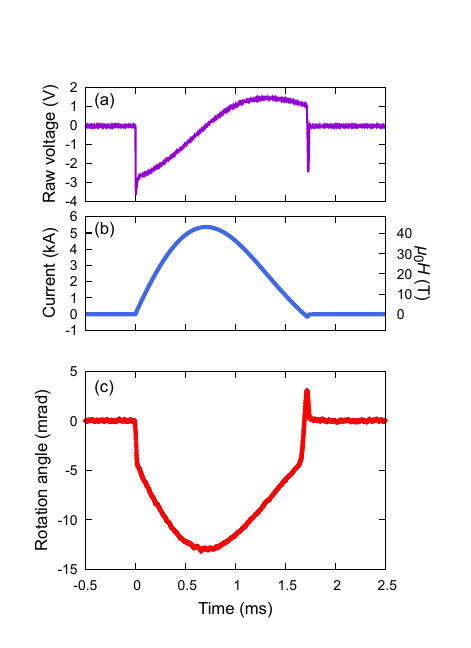}
    \caption{Time-series data of a pulse-field MOKE experiment over 40-T. The field was generated using the 6-mm coil placed in liquid nitrogen. 
    (a) Voltage signal \rev{from} the current sensor, recorded by the oscilloscope. 
    (b) Current \rev{calculated by numerically integrating} the sensor voltage. \rev{The corresponding} field  obtained \rev{using} a calibrated field-current ratio is shown \rev{on} the right vertical axis. (c) Time dependence of \rev{the} magneto-optical rotation signal (including both the sample Kerr rotation and background contribution) measured with Fe\sub{3}O\sub{4} using Eq.~\eqref{eq:Kerr_conventional}.}
    \label{fig: pulse-40T}
\end{figure}

In order to \rev{demonstrate the} applicability of our MOKE technique \rev{under} more extreme conditions, we performed measurements at 77~K on the (001) surface of a magnetite (Fe\sub{3}O\sub{4}) single crystal (MTI Corporation).
Here, the pulse-field coil \rev{was} immersed in liquid nitrogen so that we \rev{could} apply higher current due to lower lead resistance and higher cooling capacity. 
\rev{Note} that the sample is also \rev{placed} directly in liquid nitrogen, whereas the reflection is not \rev{significantly} affected even in liquid nitrogen owing to the high stability and short focal length of our ferrule-based fixture.
\rev{The} voltage \rev{obtained} from the current sensor coil \rev{for} a 6-mm bore magnet \rev{under} 1700~V charging is plotted in Figs.~\ref{fig: pulse-40T}(a).
For the data in Fig.~\ref{fig: pulse-40T}, the field value is \rev{calculated} from current, \rev{using a} calibration factor between current and field obtained \rev{in a prior} calibration.
\rev{As shown in (b), the field reached 43~T, which is very close to the maximum DC field achieved \rev{at} high-field facilities in the world~\cite{Hahn2019.Nature.570.496}. 
The time dependence of $\thetaT$ plotted in Fig.~\ref{fig: pulse-40T}(c) exhibits features similar to the data shown in Fig.~\ref{fig: pulse}(c): the steep changes at the beginning and at the end of the pulse correspond to the domain-alignment signal, whereas the gradual changes in the middle of the pulse arise from the background contribution.
}

\begin{figure}[t]
    \includegraphics[width=0.4\textwidth]{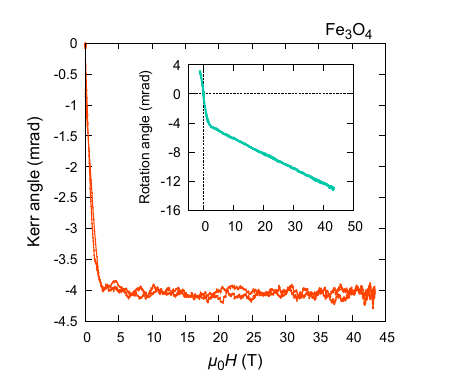}
    \caption{
    Demonstration of pulse-field MOKE measurements \rev{exceeding} 40~T on the (001) surface of magnetite (Fe\sub{3}O\sub{4}) performed at 77~K. 
    The optical power of the light source was set to \rev{approximately} 7.5~mW.
    The total MO rotation $\thetaT$, shown in the inset, is evaluated using Eq.~\eqref{eq:Kerr_conventional}.
    The main panel shows the Kerr angle $\thetaK$ of the sample obtained after subtracting the background contribution $\theta\subm{bk} = \alpha \mu_0 H$ with \rev{a} slope $\alpha = -0.205$~mrad/T.
    The saturation value ($-4.048 \pm 0.005$~mrad) is in good agreement with a \rev{previously reported} value~\cite{Fontijn1997.PhysRevB.56.5432}.
    \label{fig: 40T}
    }
\end{figure}

In Fig.~\ref{fig: 40T}, we show $\thetaT$ as a function of magnetic field. 
As shown in the inset, the raw data exhibit linear field dependence up to 43~T due to the background Faraday contribution with \rev{a} slope of $\theta\subm{bk} / (\mu_0 H) \sim -0.205$~mrad/T. 
The persistence of linear behavior indicates a successful measurement of $\thetaT$ up to 43~T.
After subtracting $\theta\subm{bk}$, we obtain $\thetaK$ of Fe\sub{3}O\sub{4}, as shown in the main panel of Fig.~\ref{fig: 40T}.
The observed saturation value is $-4.048 \pm 0.005$~mrad, which is close to the literature value on the (110) surface ($-4.36$~mrad at 0.8~eV, room temperature, and  1.76~T)~\cite{Fontijn1997.PhysRevB.56.5432} including its sign.
The small difference is attributable to anisotropy effects or influence of the charge order (Verwey transition) at around 120~K~\cite{Matsumoto1978.JPhysSocJpn.44.162}. 
The closeness of the experimental and literature values again indicates successful measurements of $\thetaK$.
The resolution of this low-temperature measurement was around 0.1~mrad, which equals the resolution achieved in the room-temperature experiment. 

To the best of our knowledge, previous reports of pulse-field MOKE are limited to \rev{only} room-temperature experiments \rev{with maximum fields of} up to 11~T~\cite{Chen2013.Measurement.46.52,Lin2017.ChinJPhys.55.698}.
Thus, the present study largely exceeds the temperature and field limits of pulse-field MOKE technique, demonstrating \rev{the} applicability of MOKE in much more extreme conditions than previously recognized.

\section{Summary}

In this work, we present a new technique \rev{for measuring} polar MOKE under pulsed magnetic \rev{fields}.
With successful measurements of MOKE above 40~T and down to 77~K with \rev{a} resolution \rev{of 0.1~mrad}, we demonstrate that MOKE is indeed a powerful technique to measure magnetic properties under pulsed \rev{fields}.
We \rev{also emphasize} that our setup is \rev{remarkably} simple and \rev{readily} applicable to stronger field apparatus \rev{at} high-field facilities, \rev{including} 1000-T destructive magnetic compression magnets~\cite{Nomura2014.PhysRevLett.112.247201, Nakamura2018.RevSciInstrum.89.095106, Matsuda2020.NatureCommun.11.3591, Ikeda2023.NatureCommun.14.1744}.
Moreover, MOKE has \rev{a key} advantage: it is applicable to small samples such as thin films or non-transparent samples including metallic materials, compared with other magnetic probes such as magnetization measurements or Faraday rotation.
Thus, for example, with pulse-field MOKE, we would be able to detect magnetic phase transitions \rev{in non-transparent, flake-like samples or thin films}, for which bulk magnetization measurements using pick-up coils or Faraday rotation measurements are not applicable.
Such studies would contribute \rev{significantly} to \rev{the} advancement of high-magnetic-field materials science.

In the end, we emphasize that the new sample fixture is \rev{also} applicable to static or zero-field experiments.
The fixture enables plug-and-play MOKE experiments free from complicated alignment \rev{procedure}, \rev{resulting in significantly} higher experimental throughput.
\rev{Thus,} this paper contributes also to \rev{the expansion and acceleration of} MOKE studies \rev{aimed at investigating} non-trivial time-reversal-symmetry breakings in various classes of materials.

\section*{Acknowledgements}

We acknowledge Y.~H.~Matsuda, Y.~Ishii, M.~Tokunaga, and H.~Mitamura for valuable comments and discussions.
We also thank \rev{Y. Hu, K.~Yada, G.~Kawabuchi, G.~Watanabe, J.~Xia, C.~Farhang, J.~Wang, I.~Kakeya, Y.~Gotoh} for technical assistance.
We acknowledge Photonic Science Technology for development of ferrule-based GI-lens focusers.
This work was supported 
by Grant-in-Aids for Academic Transformation Area Research (A) ``1000 Tesla Science’’ (KAKENHI Grant Nos.~23H04859 and 23H04861) from the Japan Society for the Promotion of Science (JSPS), 
by Grant-in-Aids for Scientific Research (KAKENHI Grant Nos.~JP24H00194, JP23K17670, JP22H01168) from JSPS,
by Japan Science and Technology Agency (JST) FOREST program (Grant No.~JPMJFR222W), 
by ISHIZUE 2023 of Kyoto University Research Development Program, 
by Iketani Science and Technology Foundation (Grant No.~0361078-A),
and by The Mitsubishi Foundation (Grant No.~202410051).
We acknowledge support for the construction of experimental setups from Research Equipment Development Support Room of the Graduate School of Science, Kyoto University; and support for liquid helium and nitrogen supplies from Low Temperature and Materials Sciences Division, Agency for Health, Safety and Environment, Kyoto University.


\bibliographystyle{./apsrev4-1_nocomma_misc-modified_with-title}
\bibliography{Kerr}

\appendix

\section{Temperature Dependence of the Reflection}
\label{sec: appendix_T-dep-reflection}

\begin{figure}
    \includegraphics[width=0.40\textwidth]{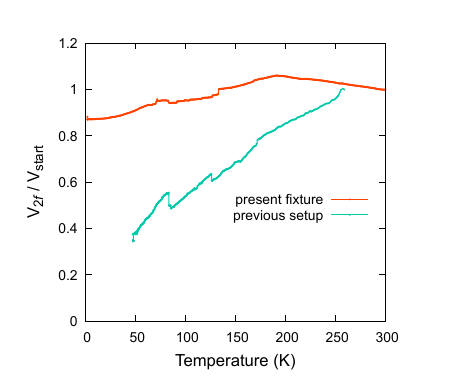}
    \caption{Example of the change in the reflection \rev{during cooling}. Here, we plot the amplitude of the 2nd harmonic signal, which is proportional to the power of the reflected light, normalized by its \rev{initial} value \rev{at the start of cooling}, $V\subm{start}$. 
    These measurements were performed in commercial ${}^4$He cryostats. 
    For the previous setup used in Ref.~\cite{Hu2023.arXiv2208.08036}, the reflected power \rev{often dropped} by several tens of percent upon cooling even if we carefully optimized the reflection at room temperature, as shown by the green curve.
    In contrast, in the present ferrule-based setup, the decrease of reflection upon cooling is at most around 10\% as shown by the orange curve. 
    We also emphasize that \rev{this ferrule-based setup eliminates the need for time-consuming} optimization \rev{before cooling}.
    }
    \label{fig: temperature}
\end{figure}

Another advantage of the present ferrule-based sample fixture is the stability of the reflection, for example, against mechanical vibration and thermal contraction.
Typical temperature dependence of the reflection is presented in Fig.~\ref{fig: temperature}.
Upon cooling down to 2~K, the intensity varies \rev{by only approximately} {\percentage{10}} with the present fixture, while the intensity often drops by more than {\percentage{50}} with our previous setup consisting of multiple screws to align the sample~\cite{Hu2023.arXiv2208.08036}.
Since the present fixture \rev{consists} of fewer parts without any moving elements, the fixture is robust against external perturbations.

\begin{figure}
    \includegraphics[width=0.39\textwidth]{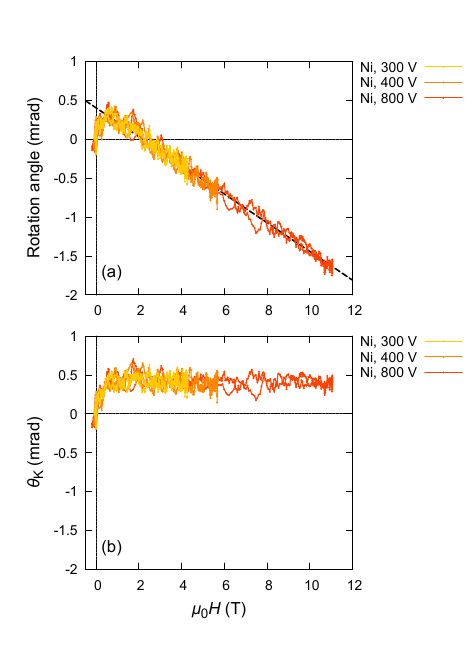}
    \caption{\rev{
    (a) Raw magneto-optical rotation signal measured with Ni. The sign change  around 2~T originates from the cancellation between $\thetaK$ and Faraday background contribution.
    The black dashed line is a fit used to evaluate the background contribution.
    (b) $\thetaK$ obtained after subtracting a linear background. 
    The saturation value is consistent with the previous reports~\cite{Buschow1983.JMagMagMat.38.1,Delin1999.PhysRevB.60.14105}.}
    }
    \label{fig: Ni_kerr-vs-field}
\end{figure}

\rev{
\section{Data for Ni}
\label{sec: appendix_Ni}

We here present graphs showing only results of MOKE measurements on Ni, in order to more clearly display its $\thetaK$, which is smaller than that of Fe. 
Figure~\ref{fig: Ni_kerr-vs-field}(a) shows the field dependence of $\thetaT$ measured using pulses of various strengths.
The sign change in $\thetaT$ observed around 2~T originates from the cancellation between $\thetaK$ and $\theta\subm{bk}$. 
After subtracting $\theta\subm{bk}$, a typical ferromagnetic field dependence of $\thetaK$ is recovered, as shown in Fig.~\ref{fig: Ni_kerr-vs-field}(b).
}

\section{Details of Numerical lock-in analysis with \rev{a} proper phase \rev{treatment}}
\label{sec: appendix_numerical-lock-in}

\subsection{Numerical lock-in analysis}

Here \rev{we begin by discussing} numerical lock-in analysis in general.
Consider a time-periodic signal of the form
\begin{align*}
V\subm{sig}(t) = V\subm{DC} + \sum_{i=0} \left[ V_{x,i} \sin(\omega_i t) + V_{y,i}\cos(\omega_i t) \right].
\end{align*}
To extract the amplitude and phase of the $\omega_j$ component, a reference signal $V_{x,{\mathrm{ref}}}(t) = V\subm{r}\sin(\omega_j t)$ is multiplied \rev{by} $V\subm{sig}$ and then is passed \rev{to} a low-pass filter (LPF) with sufficiently low cut-off frequency:
\begin{align*}
V\subm{sig}(t)\times V_{x,{\mathrm{ref}}} (t) \propto &V_{x,j}\sin^2(\omega_j t) + V_{y,j}\sin(\omega_j t)\cos(\omega_j t) \\
&+ \text{other oscillatory terms} \\
&\xrightarrow[\text{LPF}]{} \rev{\frac{1}{2}} V_{x,j}.
\end{align*}
With this process, we obtain \rev{a} DC signal proportional to $V_{x,j}$.
Similarly, multiplication by $V_{y,{\mathrm{ref}}}(t) = V\subm{r}\cos(\omega_j t)$ and subsequent use of LPF yield \rev{a} DC component proportional to  $V_{y,j}$.

This process is electrically performed in a lock-in amplifier, but \rev{it can also be implemented} numerically using time-series \rev{datasets}.
One advantage of such a numerical lock-in analysis is that types and parameters of the LPF can be \rev{adjusted} after taking the data. 
Moreover, when performing the numerical LPF operation at a certain time $t_1$, one can use both the forward and backward data.
This substantially reduces artificial response delay.
\rev{Additionally}, we can use many $n$-th harmonic signals to obtain the Kerr angle, possibly enhancing the accuracy of the obtained values (see Eq.~\eqref{eq: thetaK_new-relation}).

In this work, we \rev{employ} simple moving-average LPF with an averaging period of an integer $N$ multiple of the signal period $1/f =  2\pi/\omega$ applied \rev{symmetrically to both} forward and backward data, i.e.,
\begin{align*}
    V_{\mathrm{LPF},x}(t_1) = \frac{f}{2N}\int_{t_1-N/f}^{t_1 + N/f} V\subm{sig}(t)\times V_{x,{\mathrm{ref}}} (t) \, \mathrm{d}t.
\end{align*}
Compared with other \rev{LPFs} such as Gaussian \rev{filters}, this simple moving-avarage LPF has \rev{the} advantage \rev{of effectively rejecting higher harmonic components}~\cite{Mitamura2020.RevSciInstrum.91.125107} even \rev{with} a relatively short integration time (equivalent to time constant) $\tau \sim N/f$.
\rev{For the data shown in this paper, we use $N = 20$ for $f=1.52$~MHz, corresponding to the time constant of $N/f = 13$~$\mu$s. 
This value defines the time resolution of our experiments. 
This time resolution can be improved by selecting a smaller $N$ at the expense of the signal-to-noise ratio of the MO rotation angle.}

\subsection{Phase-resolved analysis of the photodetector signal}
\label{sec: phase-resolved-analysis}
In order to evaluate the Kerr angle including its sign, it is necessary to properly \rev{account for} the phase of the oscillatory signal.
For a measurement using the loop-less Sagnac interferometer, the photodetector signal $V\subm{PD}(t)$ under phase modulation with its depth $\phim$ and angular frequency $\omega$, with consideration of \rev{a} phase shift due to an arbitrary choice of the origin of the time $t_0$, is given by 
\begin{align*}
    V\subm{PD}(t) = V_0 + V\subm{m}\cos\left[2\phim\sin\left(\omega (t-t_0)\right) + 2\theta\subm{tot}\right],
\end{align*}
where $\theta\subm{tot} = \thetaK + \theta\subm{bk}$ is the measured total MO rotation consisting of the sample Kerr angle $\thetaK$ and the background contribution $\theta\subm{bk}$.
This \rev{expression} is rewritten using the Jacobi-Anger expansion: 
\begin{align*}
    V&\subm{PD}(t) = V_0 + V\subm{m}J_0(2\phim)\cos(2\theta\subm{tot}) \\
    & + 2V\subm{m} \sum_{n=1}^\infty J_{2n}(2\phim) \cos(2\theta\subm{tot}) \cos[2n\omega (t-t_0)] \\ 
    &- 2V\subm{m} \sum_{n=0}^\infty J_{2n+1}(2\phim) \sin(2\theta\subm{tot}) \sin [(2n+1)\omega (t-t_0)] .
\end{align*} 
This can be further rewritten in the form
\begin{align*}
V\subm{PD}(t) = V\subm{DC} + \sum_{n=1}^\infty \left[ V_{x,n} \sin(n\omega t) + V_{y,n}\cos(n\omega t) \right],
\end{align*}
where
\begin{align*}
    V_{x,n}=V_{n} \cos\delta_n, \ \ \ V_{y,n} = V_{n}\sin\delta_n.
\end{align*}
Here
\begin{align*}
    &V_{n} = 
     \begin{cases}
         2V\subm{m} J_n(2\phim)\sin(2\theta\subm{tot}) \ \ (n: \text{odd}) \\
         2V\subm{m} J_n(2\phim)\cos(2\theta\subm{tot}) \ \ (n: \text{even}) \\
     \end{cases}
\end{align*}
and
\begin{align*}
    &\delta_n = 
    \begin{cases}
         \pi - n\omega t_0 \ \ (n: \text{odd}) \\
         \pi/2 - n\omega t_0 \ \ (n: \text{even}) \\
    \end{cases}
\end{align*}
\rev{represent} the amplitude and phase of the $n$-th harmonic component, \rev{respectively,} and can be measured using \rev{a} lock-in amplifiers or by numerical lock-in analysis.

If the phases $\delta_n$ are properly evaluated, then 
\begin{align}
    &\alpha_n = 
    \begin{cases}
         \delta_n - \pi H[J_n(2\phim)] \ \ (n: \text{odd}) \\
         \delta_n - \pi H[J_n(2\phim)] + \pi/2 \ \ (n: \text{even}) \\
    \end{cases}
    \label{eq:alpha-definition}
\end{align}
\rev{should be} equal to $-n\omega t_0$, and thus $\alpha_n$ should be proportional to $n$.
Here, $H(x)$ is the Heaviside step function ($H(x) = 1$ for $x>0$ and 0 for $x<0$), which is used to correct an additional $\pi$ phase shift \rev{that occurs} when $J_n(2\phim)$ is negative.
The relation $\alpha_n = -n\omega t_0$ is used to \rev{verify the correctness} of the phase evaluation, as well as to obtain $\omega t_0$ from the slope of an $\alpha_n$-$n$ plot (For a concrete example, see Fig.~\ref{fig: phase-vs-n}).
Once $\omega t_0$ is \rev{determined}, we define the phase-correction angles
\begin{align}
    \tilde{\delta}_{n} = 
    \begin{cases}
       \pi H[J_n(2\phim)] - n\omega t_0 \ \ (n: \text{odd}) \\
       \pi H[J_n(2\phim)] - \pi/2 - n\omega t_0 \ \ (n: \text{even}) 
    \end{cases},
    \label{eq: delta-tilde-definition}
\end{align}
in order to obtain the phase corrected in-phase voltage from measured $V_{x,n}$ and $V_{y,n}$ as
\begin{align}
    V_{\mathrm{in},n} = 
        V_{x,n}\cos(\tilde{\delta}_n) + V_{y,n}\sin(\tilde{\delta}_n) 
        \label{eq: phase-corrected-in-phase-voltage}
\end{align}
which \rev{is} further used to \rev{determine} the Kerr angle in the next subsection.
Notice that it is better to use $\tilde{\delta}_n$ determined from multiple \rev{harmonic} components than to use the measured $\delta_n$, because odd-harmonic signals \rev{tend} to be very small and thus the measured $\delta_n$ for odd $n$ can contain uncertainties.

As a consistency check, it is \rev{useful} to evaluate the out-of-phase component after phase correction
\begin{align}
    V_{\mathrm{out},n} = 
         -V_{x,n}\sin(\tilde{\delta}_n) +V_{y,n}\cos(\tilde{\delta}_n).
        \label{eq: phase-corrected-out-of-phase-voltage}
\end{align}
\rev{This quantity} should be close to zero if the phase analysis is properly performed.

\subsection{Evaluation of the Kerr angle}
\label{subsec: evaluation_Kerr}

In previous MOKE studies using the loop-less Sagnac interferometer~\cite{Xia2006.PhysRevLett.97.167002,Xia2006.ApplPhysLett.89.062508, Hu2023.arXiv2208.08036}, $\theta\subm{tot}$ \rev{was} evaluated from the ratio \rev{of the} amplitudes of the 1st and 2nd harmonic signals $|V_1|$ and $|V_2|$ \rev{using the relation} $ |\theta\subm{tot}| \simeq (J_2(2\phim)/2J_1(2\phim)) (|V_1|/|V_2|).$
The modulation depth $\phim$ should be \rev{determined prior to} the MOKE measurement, by measuring \rev{the} dependence of $V_2$ on the amplitude $V\subm{m}$ of the phase modulation signal.
Typically, $2\phim = 1.84$, \rev{which corresponds to the} peak \rev{of} $J_1(2\phim)$, is used to maximize MOKE sensitivity. 
This modulation depth \rev{yields} $J_2(2\phim)/J_1(2\phim) =  0.543$ and thus we obtain $ |\theta\subm{tot}| \simeq 0.271 |V_1/V_2|$.

A very similar relation is \rev{also} applicable to the phase corrected signals:
\begin{align*}
   \theta\subm{tot} \simeq \frac{1}{2}\frac{J_2(2\phim)}{J_1(2\phim)} \frac{V_{\mathrm{in},1}}{V_{\mathrm{in},2}}
\end{align*}
The advantage of this \rev{relation} over the previous one is that \rev{it allows determination of} the sign of the MO rotation angle. 
We will demonstrate \rev{the} importance \rev{of this capacity} in Appendix.~\ref{sec: appendix_actual_process}.
For different modulation depths, this relation can be generalized \rev{as} 
\begin{align*}
   \theta\subm{tot} \simeq \frac{1}{2}\frac{J_{m\subm{e}}(2\phim)}{J_{m\subm{o}}(2\phim)} \frac{V_{\mathrm{in},m\subm{o}}}{V_{\mathrm{in},m\subm{e}}},
\end{align*}
where $m\subm{e}$ and $m\subm{o}$ are certain even and odd integers, \rev{respectively}.

We can also use another convenient relation:
\begin{align*}
    \theta\subm{tot} \simeq  \frac{V_{\mathrm{in},3}}{V_{\mathrm{in},2} + V_{\mathrm{in},4}} \sqrt{\frac{15V_{\mathrm{in},2} + 24 V_{\mathrm{in},4} + 9V_{\mathrm{in},6}}{80 V_{\mathrm{in},4}}}
\end{align*}
This equation is \rev{advantageous} because \rev{it does not require precise determination of} $\phim$ \rev{prior to} each measurement.
This advantage would be \rev{valuable} in large-scale pulse-field experiments where machine time is often \rev{severely} limited.

We here provide a \rev{derivation} of this relation.
This is based on the recurrence relation of the Bessel functions:
\begin{align*}
    J_{n+1}(x) + J_{n-1}(x) = \frac{2n}{x}J_n(x)
\end{align*}
This \rev{identity} leads to various \rev{expressions} such as $J_2(x) + J_4(x) = 6J_3(x)/x$, $J_3(x)+J_5(x) = 8J_4(x)/x$, and $J_4(x) + J_6(x) = 10J_5(x)/x$. 
From these, \rev{we} obtain
\begin{align*}
    \frac{J_2(x)+J_4(x)}{6} + \frac{J_4(x)+J_6(x)}{10}   & = \frac{J_3(x) + J_5(x)}{x} \\ 
    &= \frac{8}{x^2} J_4(x).
\end{align*}
\rev{Solving} this equation \rev{yields}
\begin{align*}
    x^2 = \frac{240 J_4(x)}{5J_2(x) + 8 J_4(x) + 3J_6(x)}
\end{align*}
This implies that the modulation depth can be obtained by measuring the amplitude of 2nd, 4th, and 6th harmonic components. 
\rev{Furthermore}, we \rev{find}
\begin{align*}
    \frac{J_3(x)}{J_2(x)+J_4(x)} = \frac{x}{6} = \sqrt{\frac{20 J_4(x)}{15J_2(x) + 24 J_4(x) + 9J_6(x)}}.
\end{align*}
\rev{Returning} to the relations $V_{\mathrm{in},n} = V\subm{m}J_n(2\phim)\cos(2\theta\subm{tot}) $ ($n$: even) and $V_{\mathrm{in},n} = V\subm{m}J_n(2\phim)\sin(2\theta\subm{tot})$ ($n$: odd), the above equation \rev{allows us to express} the \rev{MO rotation} angle from 2nd, 3rd, 4th, and 6th harmonic signals by using the relation:
\begin{align*}
    \theta\subm{tot} &\simeq \frac{V_{\mathrm{in},3}}{V_{\mathrm{in},2} + V_{\mathrm{in},4}} \frac{J_2(2\phim)+J_4(2\phim)}{2J_3(2\phim)} \nonumber \\
    & =  \frac{V_{\mathrm{in},3}}{V_{\mathrm{in},2} + V_{\mathrm{in},4}} \sqrt{\frac{15V_{\mathrm{in},2} + 24 V_{\mathrm{in},4} + 9V_{\mathrm{in},6}}{80 V_{\mathrm{in},4}}}
\end{align*}

\subsection{Actual process of the analysis}
\label{sec: appendix_actual_process}

\begin{figure}[tb]
    \includegraphics[width=0.4\textwidth]{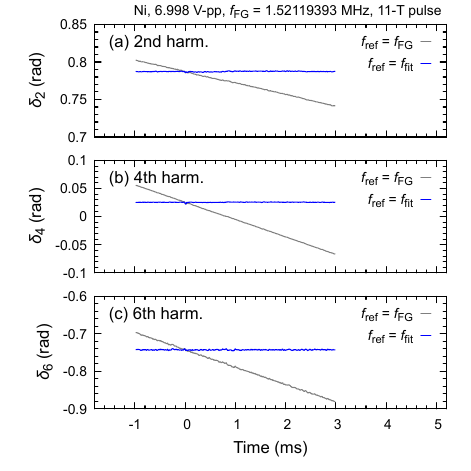}
    \caption{Comparison of the voltage-signal phase $\delta_n = \arctan (V_{y,n}/V_{x,n})$ for the harmonic components of $n=2, 4, 6$ deduced from the numerical lock-in analysis. 
    \rev{When using} the nominal function-generator output frequency $f\subm{FG} = 1.52119393$~MHz as the reference frequency $f\subm{ref}$ for the numerical lock-in analysis, artificial phase drift occurs due to a small \rev{frequency} mismatch (gray curves). By \rev{instead} using the frequency obtained from the fitting of the recorded function-generator signal, $f\subm{fit} = 1.52119270$~MHz, the drift is completely suppressed as shown with the blue curves. 
    \rev{Note} that the difference between $f\subm{FG}$ and $f\subm{fit}$ is as small as 1.23~Hz.
    The absence of the phase drift is crucial \rev{for} phase-resolved analysis.}
    \label{fig: phase-compare}
\end{figure}

In this subsection, we concretely explain the analysis process by taking the MOKE measurement of Ni foil (Nilaco, thickness : 0.05~mm; purity: 99.+\%) under \rev{an} 11-T pulse field as an example.
The frequency of the phase modulation signal output from the function generator (FG) \rev{was} determined to be $f\subm{FG} = 1.52119393$~MHz based on \rev{prior} measurements of the frequency dependence of $V_2$ using a lock-in amplifier.
The amplitude of the phase modulation \rev{was} 6.998~V-pp, corresponding to the phase modulation depth of $\phim = 2.17$~rad. 

Firstly, in order to evaluate the phase of the signal, the accuracy in the frequency of the reference signal is very important. 
There exists small but noticeable difference in the actual frequency and the nominal frequency $f\subm{FG}$. 
If we use $f\subm{FG}$ in the numerical lock-in analysis, \rev{even a slight} frequency mismatch \rev{lead to} artificial \rev{phase drift in} the lock-in voltages, as \rev{illustrated by} the gray curves in Fig.~\ref{fig: phase-compare}.
Thus, we recorded the FG signal by the oscilloscope as shown in Fig.~\ref{fig: setup} and fitted \rev{it} with a sinusoidal function to evaluate the frequency $f\subm{fit}$ with \rev{an} accuracy better than 0.01~Hz.
In the present case, $f\subm{fit}$ differs from $f\subm{FG}$ only by 1.23~Hz.
Nevertheless, \rev{this} small correction \rev{eliminates the phase drift entirely}, as shown \rev{by} the blue curves in Fig.~\ref{fig: phase-compare}.
This absence of the phase drift is \rev{crucial for performing} the following phase-resolved analysis.

\begin{figure}[t]
    \includegraphics[width=0.4\textwidth]{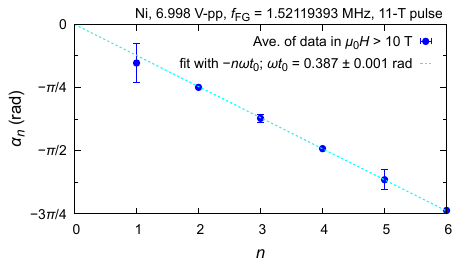}
    \caption{Phase $\alpha_n$ (defined in Eq.~\eqref{eq:alpha-definition}) of harmonic components of $n=1$ to 6. 
    Experimental values (blue points) are obtained by averaging phase \rev{data} above 10~T and the error bars \rev{indicate} the standard deviation.
    The data are well fitted by the theoretical relation $\alpha_n = -n\omega t_0$ (light-blue line), yielding $\omega t_0 = 0.387$~rad.}
    \label{fig: phase-vs-n}
\end{figure}

From the \rev{phases} of the even-order harmonic components, as well as the odd-order components under high magnetic fields, we evaluate $\alpha_n$ (\rev{defined in} Eq.~\eqref{eq:alpha-definition}), which is plotted as a function of the harmonic number $n$ in Fig.~\ref{fig: phase-vs-n}.
All data points lie on a single \rev{straight} line, \rev{consistent with the theoretical} relation $\alpha_n = -n\omega t_0$.
From the slope of this plot, we evaluate $\omega t_0$ as 0.387~rad and \rev{subsequently calculate} $\tilde{\delta}_n$ using Eq.~\eqref{eq: delta-tilde-definition}.

\begin{figure}[tb]
    \includegraphics[width=0.4\textwidth]{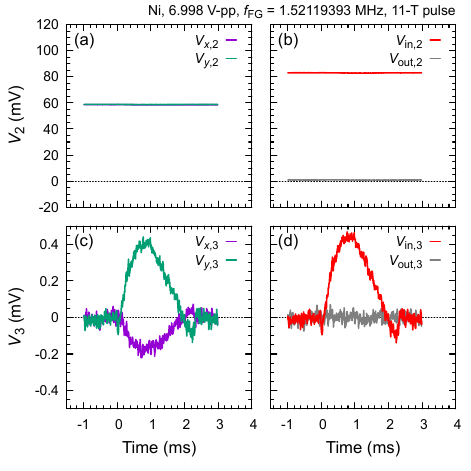}
    \caption{
    Comparison of $n$-th harmonic voltages with and without the phase correction. 
    (a) Raw voltages for the 2nd harmonic signal $V_{x,2}$ (purple curve) and $V_{y,2}$ (green curve).
    (b) In-phase and out-of-phase voltages for the 2nd harmonic signal $V_{\mathrm{in},2}$ (red curve) and $V_{\mathrm{out},2}$ (gray curve) obtained after the phase correction using Eqs.~\eqref{eq: phase-corrected-in-phase-voltage} and \eqref{eq: phase-corrected-out-of-phase-voltage} with $\tilde{\delta}_2$. 
    (c) and (d) Similar plots for the 3rd harmonic signal.
    Notice that $V\subm{out}$ is close to zero both for the 2nd and 3rd harmonic signals, indicating the validity of the phase correction analysis.
    \rev{It should also} be emphasized that the in-phase signal $V\subm{in}$ \rev{preserves} the \rev{sign information}, in contrast to the absolute value $|V_n|$.}
    \label{fig: Vx_Vy_phase-correction}
\end{figure}

We finally obtain the phase-corrected in-phase voltage $V_{\mathrm{in},n}$ using Eq.~\eqref{eq: phase-corrected-in-phase-voltage}, as exemplified in Fig.~\ref{fig: Vx_Vy_phase-correction}. 
Notice that, although $V_{\mathrm{in},n}$ resembles the absolute value $|V_n|$, the former has an important advantage that it \rev{preserves the sign information}. 
Therefore, this phase-resolved analysis is \rev{essential for evaluating} the sign of the MOKE signal.
It is \rev{also} worth mentioning that the out-of-phase component $V_{\mathrm{out},n}$ is close to zero for all $n$, as \rev{predicted by} the theory, \rev{further} supporting the validity of the analysis.

\begin{figure}[tb]
    \includegraphics[width=0.4\textwidth]{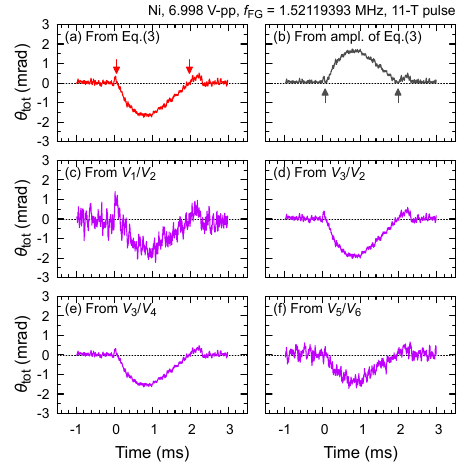}
    \caption{
    Comparison of the raw polarization-rotation angle $\thetaT$ evaluated \rev{using} various equations. 
    (a) $\thetaT$ obtained using Eq.~\eqref{eq: thetaK_new-relation}. (b) \rev{Incorrect} evaluation of $\thetaT$ using Eq.~\eqref{eq: thetaK_new-relation} but \rev{based only on} the amplitude information. 
    As indicated by the arrows in (a) and (b), the sign change occurring due to the competition between sample and background signals observed in (a) is lost in (b), demonstrating \rev{the} importance of the phase-resolved analysis.
    (c-f) $\thetaT$ from ratios (c) $V_1/V_2$, (d) $V_3/V_2$, (e) $V_3/V_4$, and (f) $V_5/V_6$. 
    The new method (a) is consistent with these conventional \rev{evaluations} (c-f).
    }
    \label{fig: Ni_Kerr-vs-time}
\end{figure}

We compare the total MO rotation angle $\thetaT$ evaluated using various relations in Fig.~\ref{fig: Ni_Kerr-vs-time}.
The result obtained from Eq.~\eqref{eq: thetaK_new-relation} with phase-resolved analysis is shown in Fig.~\ref{fig: Ni_Kerr-vs-time}(a). 
For comparison, we plot \rev{an incorrect} result obtained from the same equation but only from the amplitude information in Fig.~\ref{fig: Ni_Kerr-vs-time}(b).
As indicated by the arrows in (a) and (b), the sign changes occurring at 0.06 and 2.00~ms, \rev{originating} from a cancellation between $\thetaK$ and $\thetaT$, are replaced by extrinsic dips in the analysis without the phase consideration.
This result \rev{highlights} the importance of phase-resolved analysis explained in Sec.~\ref{sec: phase-resolved-analysis}.
In Figs.~\ref{fig: Ni_Kerr-vs-time}(c)-(d), we plot $\thetaT$ evaluated by simple ratios between odd and even harmonic in-phase signals (Eq.~\eqref{eq:Kerr_fraction_genral}).
In these evaluations, we use the nominal modulation depth $\phim = 2.1$~rad determined from modulation depth sweep measurements.
The result obtained by Eq.~\eqref{eq: thetaK_new-relation} agrees well with \rev{those from} the simple evaluations, manifesting the validity of the new relation.
Since the modulation depth is optimized for the use of $V_3$, the noise \rev{levels} in (c) and (f) are relatively \rev{high}.


\end{document}